\documentclass[english,amsmath,amssym,aps,prb,superscriptaddress,showpacs,showkeys,twocolumn]{revtex4}
\usepackage{amsmath,bm}
\usepackage{mathrsfs}
\usepackage{amsfonts}
\usepackage{graphicx}
\usepackage{color}
\usepackage{dcolumn}

\begin{document}

\title{Electronic and magnetic properties of molecule-metal interfaces: transition metal phthalocyanines adsorbed on Ag(100)}

\author{A. Mugarza}
\affiliation{Catalan Institute of Nanotecnology (ICN), UAB Campus, E-08193 Bellaterra, Spain}
\affiliation{Centre d'Investigaci\`{o} en Nanoci\`{e}ncia i
Nanotecnologia, CIN2, (ICN-CSIC), UAB Campus, E-08193 Bellaterra, Spain}

\author{R. Robles}
\affiliation{Centre d'Investigaci\`{o} en Nanoci\`{e}ncia i
Nanotecnologia, CIN2, (ICN-CSIC), UAB Campus, E-08193 Bellaterra, Spain}

\author{C. Krull}
\affiliation{Catalan Institute of Nanotecnology (ICN), UAB Campus, E-08193 Bellaterra, Spain}
\affiliation{Centre d'Investigaci\`{o} en Nanoci\`{e}ncia i
Nanotecnologia, CIN2, (ICN-CSIC), UAB Campus, E-08193 Bellaterra, Spain}

\author{R. Korytar}
\affiliation{Centre d'Investigaci\`{o} en Nanoci\`{e}ncia i
Nanotecnologia, CIN2, (ICN-CSIC), UAB Campus, E-08193 Bellaterra, Spain}

\author{N. Lorente}
\affiliation{Centre d'Investigaci\`{o} en Nanoci\`{e}ncia i
Nanotecnologia, CIN2, (ICN-CSIC), UAB Campus, E-08193 Bellaterra, Spain}

\author{P. Gambardella}
\affiliation{Catalan Institute of Nanotecnology (ICN), UAB Campus, E-08193 Bellaterra, Spain}
\affiliation{Centre d'Investigaci\`{o} en Nanoci\`{e}ncia i
Nanotecnologia, CIN2, (ICN-CSIC), UAB Campus, E-08193 Bellaterra, Spain}
\affiliation{Instituci\`{o} Catalana de Recerca i Estudis Avancats (ICREA), E-08193 Barcelona, Spain}
\affiliation{Departament de F\'{i}sica, UAB Campus, E-08193 Barcelona, Spain}

\date{\today}

\begin{abstract}

We present a systematic investigation of molecule-metal
interactions for transition-metal phthalocyanines (TMPc, with TM =
Fe, Co, Ni, Cu) adsorbed on Ag(100). Scanning tunneling
spectroscopy and density functional theory provide insight into
the charge transfer and hybridization mechanisms of TMPc as a
function of increasing occupancy of the 3$d$ metal states. We show
that all four TMPc receive approximately one electron from the
substrate. Charge transfer occurs from the substrate to the molecules, inducing a charge reorganization in FePc and CoPc, while adding one electron to ligand $\pi$-orbitals in NiPc
and CuPc. This has opposite consequences on the molecular magnetic
moment: in FePc and CoPc the interaction with the substrate tends
to reduce the TM spin, whereas in NiPc and CuPc an additional spin
is induced on the aromatic Pc ligand, leaving the TM spin
unperturbed.  In CuPc, the presence of both TM and ligand spins
leads to a triplet ground state arising from intramolecular
exchange coupling between $d$ and $\pi$ electrons. In FePc and CoPc the magnetic moment of C and N atoms is antiparallel to that of the TM. The different
character and symmetry of the frontier orbitals in the TMPc series
leads to varying degrees of hybridization and correlation effects,
ranging from the mixed-valence (FePc, CoPc) to the Kondo regime
(NiPc, CuPc). Coherent coupling between Kondo and inelastic
excitations induces finite-bias Kondo resonances involving
vibrational transitions in both NiPc and CuPc and triplet-singlet
transitions in CuPc.

\end{abstract}


\maketitle

\section{INTRODUCTION}
The interaction of molecules with solid surfaces governs the
self-assembly of supramolecular layers~\cite{Barth10NL,Gambardella10PRL} as well as the
electronic and magnetic properties of metal-organic
heterostructures~\cite{Gambardella11PRB,Gambardella10PRL} and single molecule devices.~\cite{Ratner03PRB,Bjornholm09NN} The large variety of organic and metal-organic complexes,
combined with the possibility of tuning the chemical reactivity
and electronic ground state of different complexes within the same
family of molecules, make such systems increasingly attractive for
applications. However, electronic interactions at the interface
with a metal may induce charge transfer, distort the ligand field,
and reduce electron-electron correlation effects via screening and
hybridization. The intricate interplay between all these processes
is far from being understood.

Transition-metal phthalocyanines (TMPc) represent a well-known
class of molecules with organic semiconducting properties in the
bulk and a broad spectrum of applications that includes
field-effect-transistors, gas sensors, and photovoltaic cells.~\cite{Zhang10} Due to their relatively simple and robust structure as well
as versatile chemistry, TMPc have assumed the role of a model
system to study the interaction of metal-organic complexes with
metal surfaces. Their flat adsorption geometry facilitates the
bonding of both the central TM ion and organic ligands to the
substrate, whereas their capability to coordinate many different
metal atoms allows for a systematic investigation of their
magnetic properties.

The interaction of TMPc with metal surfaces has been studied by many
techniques, including scanning tunneling microscopy (STM),~\cite{Gambardella11NC,Gambardella10PRL,Wiesendanger08PRB,Crescenzi95JoESaRP,Hofer09AC,Sakura95SS,Lyding02SS,Ortega10TJoCP,Gao07PRL,Kawai09PRL,Kawai11PRL,Tang10TJoCP,Tada05U,Zhu05S,Hou10AoCR,Bucher10TJoPCL,Bucher08PRL,Wiesendanger10PRL, Xue07PRL,Xue08PRL}  
ultraviolet and X-ray photoelectron spectroscopy (UPS-XPS),~\cite{Betti10PRB,Crescenzi95JoESaRP,Mariani09TJoCP,Stefani08TJoPCC,Fink02SS}
x-ray absorption spectroscopy (XAS),~\cite{Gambardella11PRB,Crescenzi95JoESaRP,Chasse10TJoPCL,Rossi11PRB,Gambardella10PRB} and density functional
theory (DFT).~\cite{Gao11PRB,Alouani10PRB,Dhanak10PRB,Hou08TJoPCC} Investigations
of the magnetic properties of TMPc have mostly focused on MnPc~\cite{Xue07PRL,Hou08TJoPCC},
FePc,~\cite{Gambardella11PRB,Gao07PRL,Kawai09PRL,Hou08TJoPCC} and CoPc,~\cite{Alouani10PRB,Hou08TJoPCC,Dhanak10PRB,Gambardella11PRB,Zhu05S,Hou10AoCR,Bucher08PRL,Bucher10TJoPCL,Wiesendanger10PRL}
for which it was shown that adsorption on a metallic surface tends
to quench the magnetic moment of the TM ion. A corollary to such
investigations is the implicit assumption that the magnetic
properties of TMPc at surfaces depend almost exclusively on the
ground state of the TM ion. However, a recent investigation by our
group has highlighted the possibility of inducing a magnetic
moment delocalized over the Pc organic ligand, which affects the
spin degeneracy and symmetry of the molecular ground state as well
as the electrical conductance measured at different sites within
the same molecule.~\cite{Gambardella11NC} Unpaired spins induced by
charge transfer in Pc ligands have also been found to affect the
conductance properties of TbPc$_2$ single molecule magnets
adsorbed on Au~\cite{Yamashita11NC} and their coupling to
ferromagnetic surfaces.~\cite{Gambardella11PRL} Indeed, a growing
number of experiments indicates that the organic ligand directly
affects the magnetism and transport properties of metal-organic~\cite{Bjornholm10NL,Wiesendanger10PRL,Hla10PRL}
as well as purely
organic complexes~\cite{Pascual08PRL}adsorbed on surfaces.

These results evidence the need to take into account the full
molecular structure in order to understand and predict the
magnetic properties of adsorbed systems. Here we present a
comprehensive study of four different TMPc (TM = Fe, Co, Ni, Cu)
deposited on Ag(100) in order to shed light on the mechanisms that
lead to changes of both the TM and ligand magnetic moments at the
TMPc/metal interface. Scanning tunneling spectroscopy (STS)
combined with DFT allows us to elucidate the role of charge
transfer, hybridization, and correlation at the molecule-metal
interface on the magnetic properties. The paper is organized as
follows: after presenting the experimental and theoretical methods
(Section~\ref{sec:Methods}), we first introduce the theoretical
electronic structure of TMPc in the gas-phase
(Section~\ref{sec:Pristine MPc}), which serves as the starting
point of our investigation. We then describe the experimental and
theoretical adsorption geometry of the molecules on Ag(100)
(Section~\ref{sec:adsorption configuration}), including a
discussion of electronic chirality effects, which were revealed in
a previous work on CuPc.~\cite{Gambardella10PRL} The electronic structure
of adsorbed TMPc is discussed in Section~\ref{sec:elec_struc},
where STS data are compared with the molecular projected density
of states (PDOS) derived from ab-initio DFT calculations. The
magnetic properties of TMPc are addressed in
Section~\ref{sec:mag_struc} using STS to measure the intensity
and spatial extension of Kondo anomalies in the differential
conductance, including the coupling to vibrational and magnetic
degrees of freedom in NiPc and CuPc as well as the theoretical
description of triplet-singlet excitations in CuPc using a
multi-orbital approach. Further, we calculate the magnetic moments
of adsorbed TMPc and compare them to the gas-phase magnetic
moments in neutral and negatively charged species in order to
understand the effects of charge transfer and intramolecular
exchange and their dependence on the TM ion. Finally,
Section~\ref{sec:Conclusions} summarizes the more general
conclusions of this work.

\section{METHODS}
\label{sec:Methods}
\subsection{Experiment}

TMPc were evaporated in ultra high vacuum (UHV) from a heated
quartz crucible onto a sputter-annealed Ag(100) single crystal
kept at room temperature, after degassing the 99$\%$ pure powder
material (Sigma Aldrich) for 24 hours. The deposition rate was
$\sim$ 0.05 monolayers/min; the base pressure during evaporation
was below $5 \times 10^{-10}$~mbar. Spectroscopic measurements
were performed using an STM operating at 5~K. Conductance
($dI/dV$) spectra were obtained with the lock-in technique, using
a bias voltage modulation of frequency 3 kHz and amplitude
1~mV$_{rms}$ for the low energy spectra ($-0.1<V<0.1$~V), and
3~mV$_{rms}$ for the larger energy range ($-2<V<1$~V). A
background spectrum acquired on the bare Ag surface with the same
feedback conditions was subtracted to all spectra in order to
enhance molecular resonances and minimize features originating
from the tip or substrate electronic structure.~\cite{Kern08RoSI}
Hence, only relative $dI/dV$ values are significant in these spectra. $dI/dV$
maps were acquired in the constant current mode, with an amplitude
of 3~mV$_{rms}$. Maps of the Kondo resonance were obtained using
the $d^2I/dV^2$ signal in order to scan at a finite
voltage.~\cite{Gambardella11NC} Both topographic and conductance images
were processed using the WSxM software.~\cite{Baro07RoSI} Further
details on the spectroscopic methods can be found in the
Supplementary Information of Ref.\onlinecite{Gambardella11NC}.

\subsection{Theory}

The theoretical electronic structure of both gas-phase and
adsorbed TMPc have been obtained from {\it ab-initio} calculations
using the VASP implementation of DFT in the projected augmented
plane wave scheme.~\cite{Joubert99PRB,Furthmuller96CMS} Different
approximations, namely LDA,~\cite{Zunger81PRB} GGA,~\cite{Ernzerhof96PRL}
GGA+vdW,~\cite{Grimme06JoCC,Angyan10TJoPCA} and GGA+U~\cite{Sutton98PRB} have been
compared in order to obtain consistent results, and to check the
effect of different exchange-correlation approximations or the vdW
interaction. Unless otherwise stated, the results presented in the
next sections were obtained using GGA+U~\cite{Sutton98PRB} and vdW
corrections using the scheme of Grimme.~\cite{Grimme06JoCC,Angyan10TJoPCA}
U$_{eff}$=U-J was chosen to be 3~eV in all cases. Additional tests
were made to check that reasonable values of U$_{eff}$ do not
qualitative change our conclusions. The planewave cutoff energy
was set to 300 eV. The calculated slab included 5 Ag atomic layers
intercalated by 7 vacuum layers in the vertical direction, and a
7$\times$7 lateral supercell. The positions of all atoms in the
molecule and the first three Ag layers were relaxed vertically and
laterally until forces were smaller than 0.05 eV/\AA. The
projected density of states (PDOS) of NiPc have been used to
compare single (1$\times$1$\times$1) and multiple
(5$\times$5$\times$1) $k$ point calculations, which converge after
applying a broadening of 100 meV to the data. Based on that, the
results for a single $k$ point with the latter broadening are used
in the following. Charge transfer and local magnetic moments have
been calculated using a Bader charge analysis.~\cite{Cade67TJoCP,Henkelman09JoPCM}

\section{GAS-PHASE TMPc}
\label{sec:Pristine MPc}

\begin{figure*}[!t]
\includegraphics[width=1.9\columnwidth]{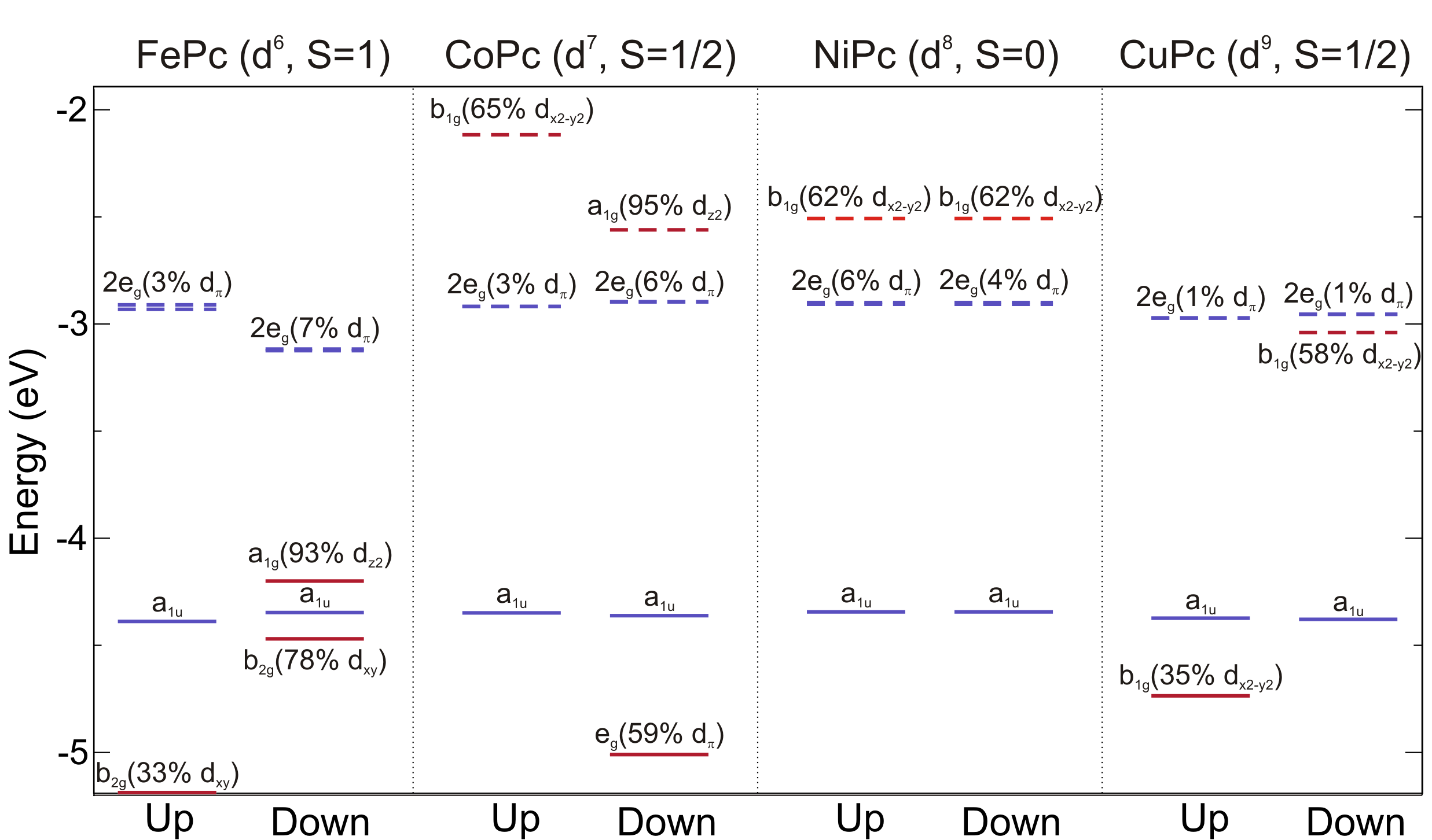}
\caption{(Color online). Spin-polarized electronic structure of
gas-phase TMPc. (Un-) occupied states are represented by (dashed)
full lines. The contribution of different $d$-states is indicated
in percentage. \label{FIG_1}}
\end{figure*}

Before discussing the effects of adsorption on the molecular
electronic and magnetic structure, we briefly introduce the
chemical configuration of gas-phase TMPc, as well as the
electronic structure obtained from spin-polarized calculations,
which will be used later as a reference.

TMPc are characterized by a rather simple and stable molecular
structure, where the TM ion bonds to four isoindole (pyrrole +
benzene) ligands via their pyrrole N (N$_p$). That leaves the ion in a
[TM]$^{2+}$ state, and the molecule with the square planar
$D_{4h}$ symmetry. Under this symmetry group, the $d$-states
transform as  $b_{2g}$ ($d_{xy}$), $b_{1g}$ ($d_{x^2-y^2}$),
$a_{1g}$ ($d_{z^2}$), and $e_g$ ($d_{xz}$,$d_{yz}=d_\pi$).
Depending on their symmetry and energy position, these orbitals
mix to a different degree with 2$p$-states of the C and N atoms.
The highest occupied and lowest unoccupied molecular orbitals
(HOMO and LUMO) of the Pc ring are represented by delocalized
$a_{1u}$ and $2e_g$ $\pi$-orbitals, respectively, with marginal
contribution from the TM $d$-states. The TM-related MO can be
classified in two groups according to their parallel ($b_{2g}$,
$b_{1g}$ = $d_\parallel$) or perpendicular ($a_{1g}$, $e_g$ =
$d_\perp$) orientation with respect to the molecular plane, and
hence, the substrate.

Figure~\ref{FIG_1} shows the spin-up and spin-down
energy levels of TMPc near the frontier orbitals obtained within
the GGA+U scheme. The 3$d$ contribution in each MO is indicated
between parentheses; the molecular magnetic moments are discussed
in Sect. \ref{sec:mag_struc}. Our calculations show that the
electronic structure of the Pc ring is barely affected by the TM
ion. This is clearly seen in the $a_{1u}$ and $2e_g$ ligand
orbitals: the $a_{1u}$ MO, with negligible $d$-contribution,
appears at the same energy in all molecules. The $2e_g$ energy
varies slightly with the $d$ configuration of the TM ion as this
orbital exhibits a finite $d_\pi$ contribution that ranges from
7$\%$ in FePc to 1$\%$ in CuPc.

The evolution of the electronic configuration of the central
[TM]$^{2+}$ ions cannot be rationalized by the simple filling of
one-electron energy levels split by the ligand field, due to the
strong $d$-$d$ correlation and the energy-dependent hybridization
with ligand states. The situation is most critical for the ground
state of FePc, where the quasidegeneracy between different
electronic terms~\cite{Gambardella11PRB,Butler88CPL} has led to a
long-standing debate in literature.~\cite{Thorp68JCP,Figgis91IC,Scheiner01JoCP,Oison09PRB,Kronik09APA,Knupfer10EPJB}
The discussion of such differences is beyond the scope of this
paper. However, the spin moment that we obtain for this molecule
is S=1, independently of the DFT functional and in good agreement
with previous
work.~\cite{Gambardella11PRB,Butler88CPL,Thorp68JCP,Figgis91IC,Scheiner01JoCP,Oison09PRB,Kronik09APA,Knupfer10EPJB}
CoPc, with one electron more, presents a robust $^2A_{1g}$ (S=1/2)
ground state with a single $a_{1g}$ ($d_{z^2}$) hole, in agreement
with x-ray absorption measurements.~\cite{Gambardella11PRB} NiPc,
with $d^8$, has a singlet (S=0) $^1A_{1g}$ ground state. The TM
spin is recovered in CuPc, with a single hole in $b_{1g}$
($d_{x^2-y^2}$), resulting in a doublet (S=1/2) $^2B_{1g}$ ground
state.

\section{ADSORPTION OF TMPc ON Ag(100)}
\label{sec:adsorption configuration}
\subsection{Adsorption geometry}
\begin{figure}[t]
\includegraphics[width=\columnwidth]{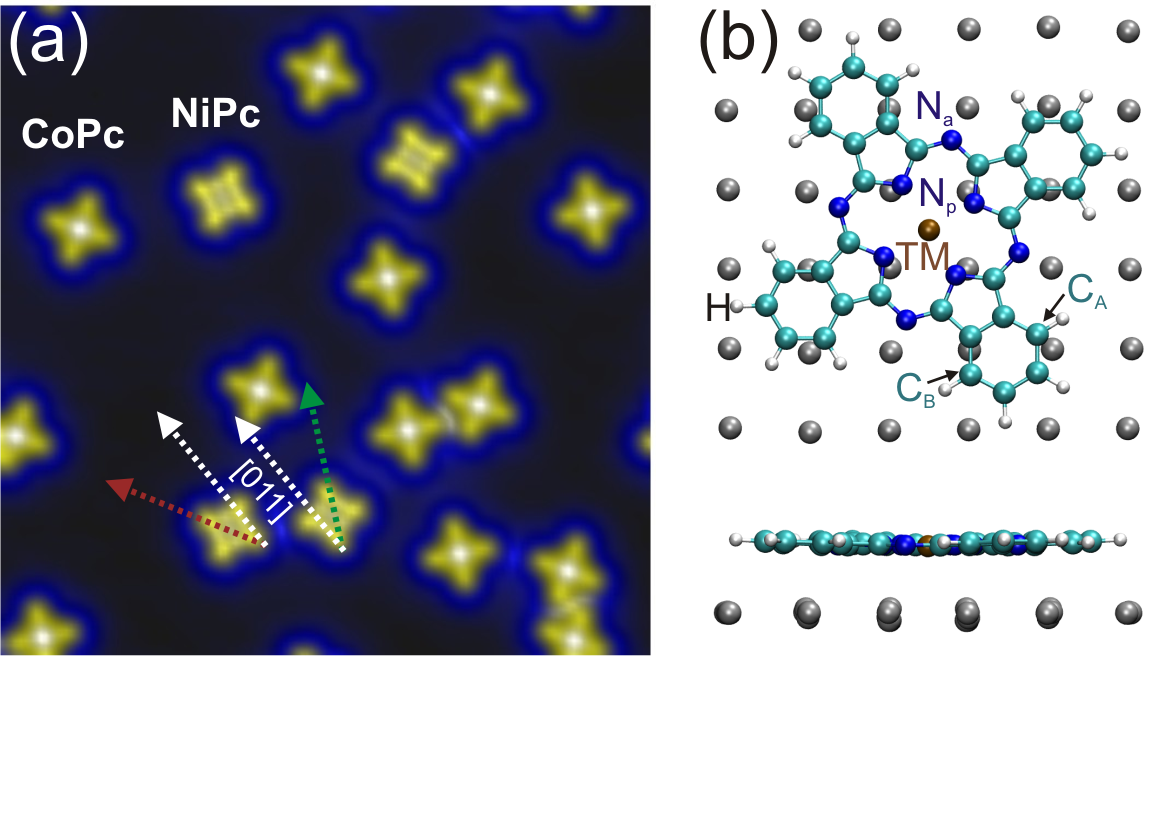}
\caption{(Color online). a) Topographic image of CoPc and NiPc
molecules co-deposited at room temperature on the Ag(100) surface
($I$ = 0.49 nA, $V$ = -1.0 V), image size $13.4 \times
13.4$~nm$^2$. The green/red arrow indicates the $+/-30^\circ$
rotation of the molecular axis with respect to the surface lattice
vectors (white arrows). b) Adsorption geometry of NiPc on Ag(100)
calculated by DFT. The different elements are labeled. N$_1$ and N$_2$ refers to
aza and pyrrole N, and C$_{A}$ and C$_{B}$ to the two C atoms of
the benzene ring used to measure chiral distortions (Table
\ref{tab1}). \label{FIG_2}}
\end{figure}

The adsorption of individual TMPc has been studied by STM. Their
topographic appearance already reflects differences that allow us
to classify them in two groups: the TM ion appears as a protrusion
in FePc and CoPc and a depression in NiPc and CuPc, as shown in
Figs.~\ref{FIG_2}~(a) and \ref{FIG_3}. The different
contrast is due to the large (small) coupling of the TM $d_\perp$
($d_\parallel$) orbitals near the Fermi level ($E_F$) with the tip
states.~\cite{Mazur96JotACS} As typical for TMPc on metals, the
molecules adsorb with the aromatic plane parallel to the surface.
Atomically resolved images of the surface reveal two different
azimuthal orientations of the TM-N$_p$ ligand axis [Fig.
\ref{FIG_2} (b)], corresponding to $\pm 30^\circ$ with respect
to the (100) surface lattice vectors. Previous systematic {\it
ab-initio} calculations of CuPc on Ag(100) revealed that this
orientation, with the TM ion located on the Ag hollow site,
corresponds to the minimum of the adsorption energy.~\cite{Gambardella10PRL} A similar adsorption site has been found
for CuPc on Cu(100) (Ref.~\onlinecite{Chiang89PRL}). The rotation
of the Pc macrocycle with respect to the substrate high symmetry
directions was attributed to bond optimization between aza-N
(N$_a$) and Ag atoms. This study confirms the leading role of the
N$_a$-Ag interaction, as both theory and experiment find very
similar adsorption configurations for all TMPc, independently of
the central TM ion (see Fig.~\ref{FIG_2}b).

The calculations also show that TMPc become slightly concave after
deposition, with the periphery of the molecule slightly further from the surface. This effect is most pronounced in the case of FePc.
The molecules also induce a small distortion of the substrate,
pushing the Ag atoms below N$_p$ slightly below the
surface plane and the Ag atoms below N$_a$ slightly above it.
The molecule-substrate distances, calculated as the difference in
the $z$ coordinate between the TM atom and the Ag atom below N$_p$, are shown in Table~\ref{tab1}. The distance obtained
with LDA is $\sim$2.5 \AA\ for the four molecules. This method is
known to overbind, compensating for the absence of van der Waals
(vdW) forces. These forces are included in the GGA+vdW method. The
results with GGA+vdW show that $z$ increases to $\sim$2.7 \AA,
keeping it TM-independent. Similar values were also found using
both methods for CoPc adsorbed on Cu(111) (Ref.
\onlinecite{Alouani10PRB}). On the less reactive Au(111) surface,
however, distances obtained with LDA for different TMPc do not
level out and still reflect a TM-dependent
behavior.~\cite{Gao11PRB,Hou08TJoPCC} In our case, we can separate the
effect of the interaction between the TM ion and the substrate by
using plain GGA, i.e., by switching off the vdW interaction. Using
this method, we find that FePc and CoPc, with $z=2.76$ and 3.08
\AA\, respectively, are significantly closer to the surface than
NiPc and CuPc, with $z=3.59$ and 3.47 \AA\, respectively. The
stronger TM-Ag interaction of FePc and CoPc is attributed to the
direct participation of the $d$-orbitals to the molecule-substrate
bonds.~\cite{Gao11PRB,Hou08TJoPCC,Betti10PRB} The vdW interaction introduced
by GGA+vdW can be estimated by the correction added to the total
energy calculated by GGA alone. This correction is larger by about
1~eV for NiPc and CuPc compared to FePc, which compensates for the
lack of direct TM-substrate interaction in the former species and
results in the nearly TM-independent bond length reported in
Table~\ref{tab1}.

\subsection{Electronic chirality}
\begin{figure}[]
\includegraphics[width=\columnwidth]{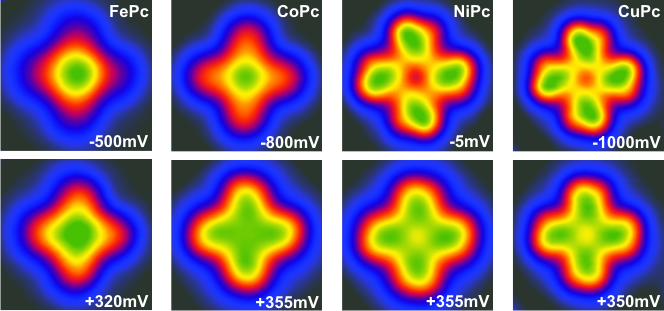}
\caption{(Color online). Bias dependent topographic images of all
TMPc. For CuPc and NiPc negative voltage images with maximum
chiral contrast are displayed. The contrast of CoPc and FePc does
not vary in the range of negative voltages investigated. For
positive voltage, all TMPc appear achiral in the bias range $V<+$1
V. \label{FIG_3}}
\end{figure}

Figure~\ref{FIG_3} shows that NiPc and CuPc present chiral
contrast at negative bias, whereas FePc and CoPc appear achiral at
both negative and positive bias. This effect has been studied in
detail for CuPc/Ag(100) in Ref.~\onlinecite{Gambardella10PRL},
where it has been shown that the chiral contrast observed by STM
is not related to nuclear distortions, but originates from the
asymmetric electronic interaction between the $a_{1u}$ and $2 e_g$
orbitals and Ag states. This asymmetry is due to the misalignment
between surface and molecular symmetry axes, where the $+30^\circ$
and $-30^\circ$ configurations imprint chiral charge distortion of
opposite sign. Here we extend this study to FePc, CoPc, and NiPc,
and quantify the contribution of conformal distortions to the
chiral contrast. We use the height difference between two opposite
C atoms at the benzene ring as a measure of their torsion (see
Table \ref{tab1} and Fig.~\ref{FIG_2}b). Its small value, close
to the calculation's accuracy, obtained for all TMPc confirms the
electronic origin of chirality. This is in line with the voltage
dependence of the chiral contrast, and the fact that it is only
observed in CuPc and NiPc in spite of the similar geometry found
for all molecules.

\begin{table}[]
\centering

\begin{tabular}{ccccc}
         \hline
        \hline
           &   GGA &  GGA+vdW &  LDA & $z$(C$_A$-C$_B$)   \\
        \hline
        FePc   & 2.76 & 2.64 & 2.43 & 0.04 \\
        CoPc   & 3.08 & 2.72 & 2.49 & 0.01 \\
        NiPc   & 3.59 & 2.74 & 2.53 & 0.02 \\
        CuPc   & 3.47 & 2.79 & 2.46 & 0.02 \\
        \hline
        \hline
\end{tabular}

\centering \caption[] {\label{tab1} Computed height distances $z$ (in \AA)
between the TM ion and the Ag atom below N$_p$, obtained within different DFT approximations, and height difference between
C$_{A}$ and C$_{B}$ atoms at the end of the benzene ring (see Fig.~\ref{FIG_2}b) obtained with GGA+vdW.}
\end{table}

Electronic chirality is imprinted to each MO to a different
degree, depending on their spatial distribution and hybridization
with the substrate, with maximum effect in orbitals exhibiting a
nodal plane on the TM-N$_p$ ligand axis (see Fig.~\ref{FIG_5}).
The $a_{1u}$ MO exhibits the strongest distortion because of its
double lobe structure at the benzene ring, where the asymmetry of
charge distortion is maximum. The chiral contrast at the $2e_g$ MO,
also with nodal planes on the ligand axis, varies with energy: at
positive bias it is strongly reduced, but becomes more intense
when shifted to negative bias due to charge transfer. The MOs with
significant TM-$d$ contribution do not have nodal planes on the
ligand axis, and hence do not present any significant chirality.
This leads to a voltage-dependent chiral appearance of the TMPc
that depends solely on the specific electronic structure of each
molecule, as shown in Fig.~\ref{FIG_3}. CuPc and NiPc are
strongly chiral at negative bias due to the asymmetry of the
$a_{1u}$ orbital and of the partially occupied $2e_g$ orbital, and
achiral at positive bias, where tunneling occurs via a more
symmetric $2e_g$ resonance. In CoPc and FePc, on the other hand, the $2e_g$ orbital only appears at positive bias, where the chiral contrast is insignificant. Furthermore, TM-$d$ states and other other interface-related resonances with maximum intensity on the ligand axis hinder the
chirality of the $a_{1u}$ state.

\section{ELECTRONIC STRUCTURE}
\label{sec:elec_struc}

In this section we present scanning tunneling spectroscopy (STS)
measurements of the molecules adsorbed on the Ag(100) surface. The
energy of the MOs and other interface-related features is obtained
from individual $dI/dV$ spectra, whereas maps of the $dI/dV$
intensity at specific energies reveals their symmetry and spatial
distribution within the molecule. Experimental results are then
compared to the theoretical density of states projected onto
different MOs, which allows us to assign the observed $dI/dV$
peaks to specific orbitals, providing insight into the formation
of hybridized molecule-substrate states, charge transfer, and
molecular magnetism.

\subsection{Scanning tunneling spectroscopy}
\label{sec:Spectroscopy}

\begin{figure*}[!t]
\includegraphics[width=1.9\columnwidth]{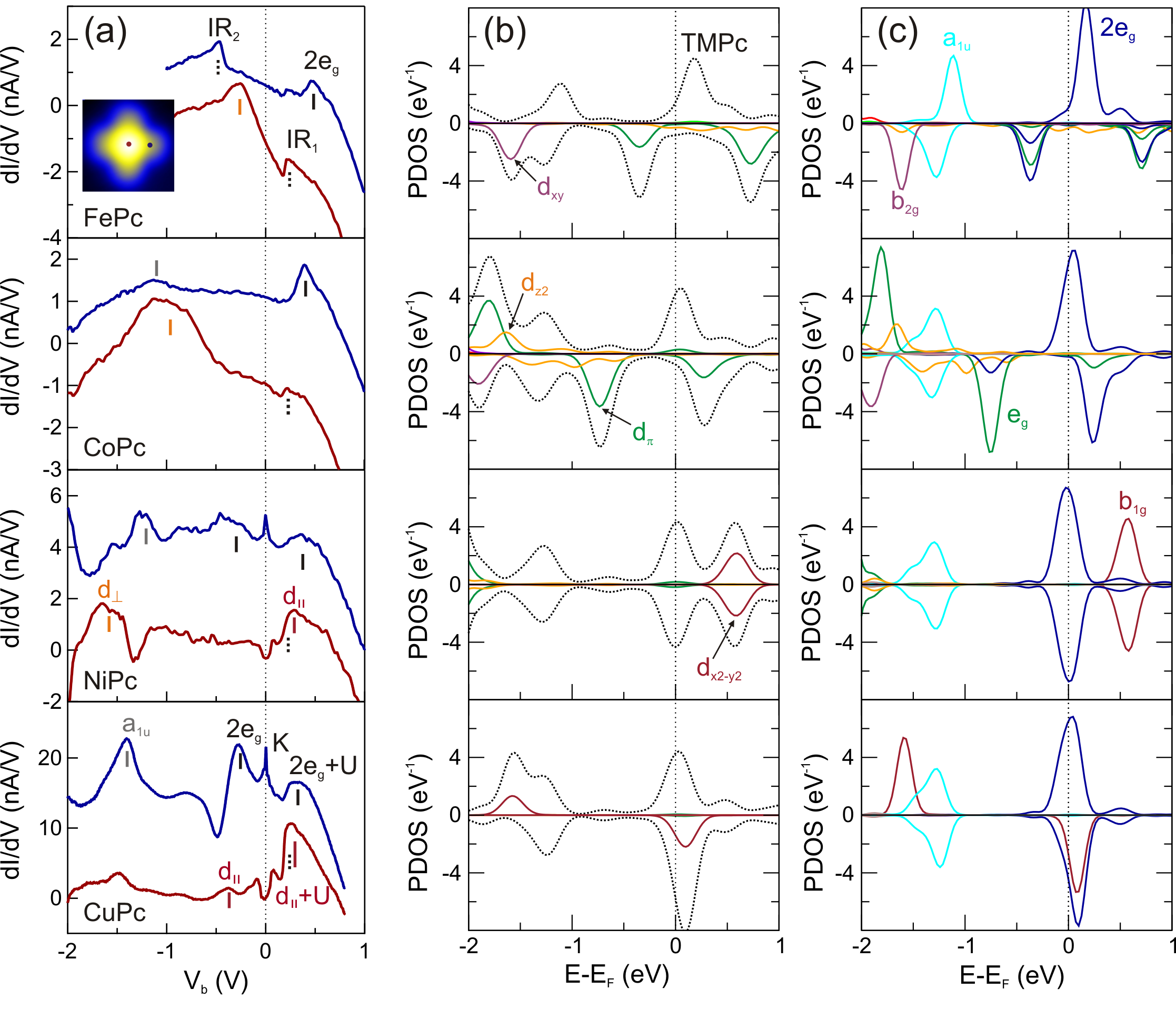}
\caption{(Color online). (a) $dI/dV$ spectra, acquired on the TM
ions (red, bottom) and benzene rings (blue, top). The latter
spectra have been offset for clarity. MOs, interface (IR) and Kondo (K) resonances, as well as Fano-type (F) features are labeled. Tunneling conditions ($I$,
$V$):  1 nA, -1.0 V (FePc, CoPc), 3 nA, -2.2 V (NiPc), 3 nA, -2.0
V (CuPc). (b) Computed spin-polarized DOS of TMPc (black), and its
projection onto the TM $d$-states. (c) Computed spin-polarized PDOS
projected onto MOs. \label{FIG_4}}
\end{figure*}

\begin{figure}[!t]
\includegraphics[width=\columnwidth]{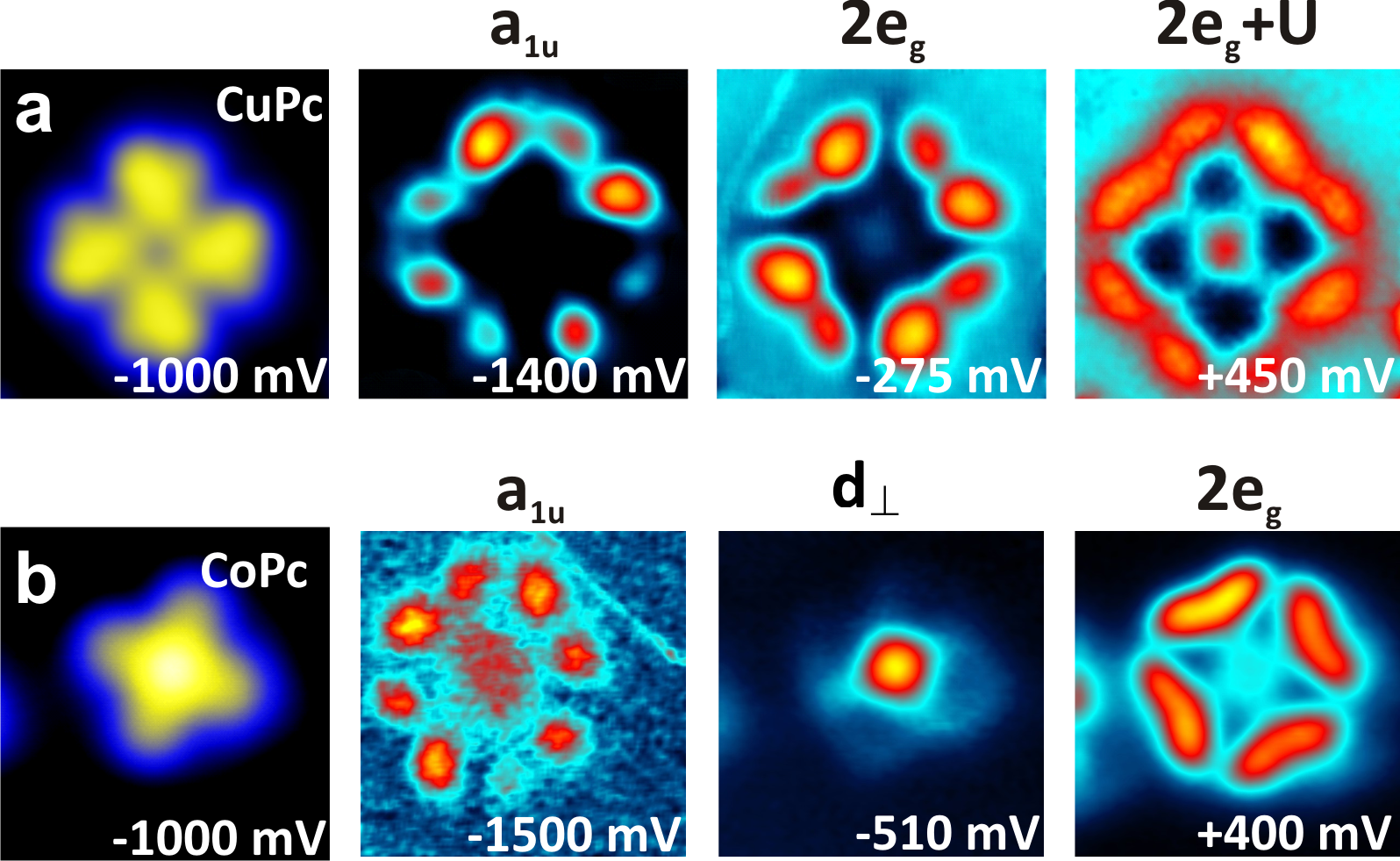}
\caption{(Color online). Constant current $dI/dV$ maps of (a) CuPc
and (b) CoPc. A topographic image of each molecule recorded
simultaneously during the acquisition of the maps is displayed on
the left side. The color scale is adjusted for maximum contrast in
each map. NiPc and FePc present similar MO distributions as CuPc
and CoPc, respectively. \label{FIG_5}}
\end{figure}

Figure \ref{FIG_4}(a) shows the $dI/dV$ spectra of FePc to
CuPc recorded by positioning the STM tip on the central TM atom
(red curves) and on a peripheral benzene ring (blue curves). The
$a_{1u}$ and $2e_g$ resonances corresponding to the HOMO and LUMO
of the gas-phase Pc ring can be easily identified by studying the
spatial distribution of the $dI/dV$ peaks of the spectra acquired
on benzene, as shown in Fig.~\ref{FIG_5}. The intensity
distribution of these resonances is similar to that found for
other substrates,~\cite{Hou10AoCR} except for the chiral
fingerprint discussed in the previous Section. The energy position
of the $a_{1u}$ state is similar in all TMPc, varying continuously
from -1.14 V in CoPc to $-1.40$ V in CuPc (FePc was not measured
in this energy range). On the other hand, the spectral
distribution of the $2e_g$ state critically depends on the TM ion:
in FePc and CoPc, a single unoccupied peak is observed at +0.39
and +0.47 V, respectively, as expected for the Pc LUMO.
Conversely, CuPc and NiPc exhibit two peaks around the Fermi
level, located at -0.29/+0.35 V and -0.35/+0.35 V, respectively.
Both peaks, however, can be associated with the $2e_g$ state due
to their similar spatial distribution. The position of the lowest
energy peak below $E_F$ indicates that this orbital is partially
occupied. The energy splitting of $\sim0.65-0.70$ eV is within the
range of the Coulomb repulsion energies obtained for
$\pi$-orbitals in aromatic complexes of similar size.~\cite{Pascual08PRL,Cohen08PRL,Anders11PRB} We therefore assign
the peaks at negative and positive bias to single and double
occupation of the $2e_g$ state, respectively, in agreement with
the results reported in Ref.~\onlinecite{Gambardella11NC}. The sharp
peak appearing between the two $2e_g$ "Coulomb blockade" peaks is
the Kondo resonance associated to the unpaired ligand
spin,~\cite{Gambardella11NC} which will be discussed further in Section
\ref{sec:mag_struc}. Note that the
short electron lifetime of empty orbitals in molecules coupled to
metals prevents their double occupation by tunneling electrons, which would lead to similar "Coulomb blockade" peaks in the $2e_g$ state of FePc and CoPc.

Spectra recorded on the TM ions, shown as red curves in Fig.
\ref{FIG_4}(a), allow us to identify the MOs with
$d$-character. The broad peak shifting downwards with increasing
$d$-occupation from FePc to NiPc can be related to $d_\perp$
states, easiest to observe due to their effective coupling with
the STM tip. In CuPc, these states are situated below the probed
energy range. The localization of this peak on the TM ion is
confirmed by the $dI/dV$ map of CoPc at -510~mV
[Fig.~\ref{FIG_5}(b)]. In this molecule, the $d_\perp$
resonance appears well below $E_F$, suggesting the complete
filling of the $a_{1g}$ ($d_{z^2}$) orbital, and hence a switch in
the charge transfer channel from the ligand $2e_g$ to TM-$d$
states compared to NiPc and CuPc. In FePc, the tail of the
$d_\perp$ peak crosses $E_F$, suggesting that these states
are not fully occupied, similarly to FePc on Au(111)
(Ref.~\onlinecite{Gambardella11PRB}).

Contrary to the $d_\perp$-states, the perturbation of the
$d_\parallel$-states due to adsorption appears to be minor.
Although confinement in the molecular plane makes them difficult
to access by STS, some features in the spectra of NiPc and CuPc
can be tentatively assigned to the $b_{1g}$ ($d_{x^2-y^2}$) state.
In the pristine molecules, this orbital is singly occupied in CuPc
and unoccupied in NiPc (see Fig.~\ref{FIG_1}). The
intensity corresponding to the unoccupied $b_{1g}$ state is masked
by a step-like interface resonance feature (IR$_1$) that we
observe around +0.2 V in all molecules. However, both CuPc and
NiPc exhibit a larger $dI/dV$ signal in this energy range compared
to CoPc and FePc, which suggests that there is an additional
orbital contributing to the tunneling current above $E_F$. The
singly occupied $b_{1g}$ state in CuPc can be assigned to the
small hump observed around -0.4~V, close to where the $2e_g$
resonance on the benzene spectra presents a pronounced valley.
This valley, absent in NiPc at this energy, is tentatively
assigned to a Fano resonance originating from the tunneling
interference between the occupied $b_{1g}$ and the more hybridized $2e_g$ states,
analogous to that occurring between resonant and direct channels in impurity and quantum dot systems.~\cite{Meirav00PRB,Wingreen98S,Maslova10SSC} Indeed, the dip observed
in NiPc at -1.4~V, where $d_\perp$ states overlap with the
$a_{1u}$ orbital, might have a similar origin, supporting the
presence of Fano peaks in these molecules.

\begin{figure}[!h]
\includegraphics[width=\columnwidth]{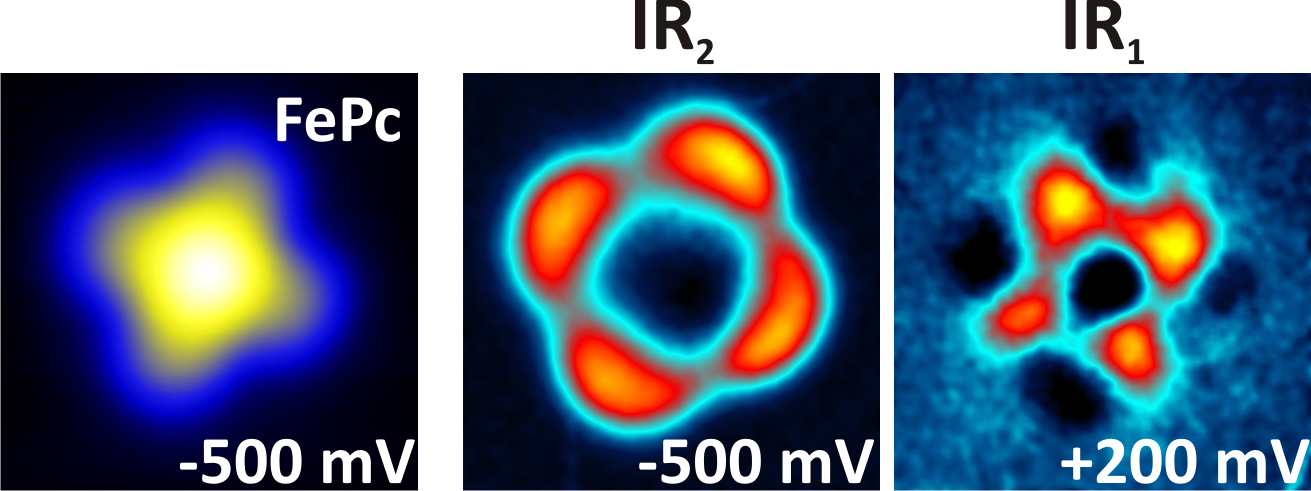}
\caption{(Color online). Constant current $dI/dV$ maps of
interfacial resonances in FePc arising from the hybridization between
molecular and substrate electrons, marked as IR in  Fig.~\ref{FIG_4}(b). The topography of FePc recorded during the
acquisition of the maps is displayed on the left side.
\label{FIG_6}}
\end{figure}

In addition to MO resonances that maintain the original character
of the TMPc states, we find additional interface resonances that
originate from strong hybridization with Ag electrons (see Fig.~\ref{FIG_6}). The IR$_1$
resonance observed at +0.2 V [Fig.~\ref{FIG_4}(a)] is
common to all TMPc and only observed on Ag(100). This resonance
could originate from the hybridization of
molecular states with the lowest lying unoccupied surface state of
Ag(100), which is close in energy.~\cite{Liu81PRL} In FePc, we observe an additional feature (IR$_2$) between the
$a_{1u}$ and $d_\perp$ states, which we identify with an interface state with intensity at
the benzene rings, similar to that reported for CoPc on Au(111)
(Ref.~\onlinecite{Hou10AoCR}).

\subsection{Density of states calculated by DFT}
\label{sec:PDOS}

In order to correlate the features observed in the $dI/dV$ spectra
with the molecular electronic structure, we have computed the PDOS projected onto the TM-$d$ states as well as on
the MOs of the TMPc, as shown in Fig.~\ref{FIG_4}. In general, we find
fair agreement between the measured and calculated resonances,
supporting the assignment of the spectral features to specific MO
based on the symmetry and spatial distribution of the $dI/dV$
intensity.

The strongest hybridized states are the $a_{1g}$ MOs of FePc and
CoPc, as expected owing to their dominant $d_{z^2}$ character. The
TM $d_{z^2}$-electrons hybridize with the Ag-$sp_z$ states,
confirming the strong, direct substrate-TM interaction
discussed in Section~\ref{sec:adsorption configuration}.
Additionally, in the case of FePc and CoPc, two $2e_g$ and one $e_g$ state
are mixed together, projecting onto one occupied and one empty
spin-down resonances (see Fig.~\ref{FIG_4}c). Mixing is induced by hybridization with the
substrate, showing that the picture of slightly distorted MO is no
longer valid in this case. This is in line with the appearance of interface resonance as those observed in STS maps (IR$_2$ in Fig.~\ref{FIG_6}). For NiPc and CuPc, hybridization with
the substrate turns out to be much smaller, and the PDOS resembles
that of the gas-phase molecules of Fig.~\ref{FIG_1}. In
these two cases the confinement of the $b_{1g}$ orbital in the
molecular plane and its $\sigma$ character reduces hybridization
with Ag electrons, diminishing the charge transfer to this
orbital. Hence, it appears unoccupied in NiPc and singly occupied
in CuPc, as in the gas-phase.

The $a_{1u}$ orbital, fixed at about -1.30 eV, is the only MO that
is not affected by the $d$-occupation of the TM ion. In the case
of FePc, the spin up and spin down $a_{1u}$ states are
exchange-split due to the single-configurational nature of DFT. This broken symmetry reflects the strong
spin-polarization of the whole molecule. The position of the
$2e_g$ MO, on the other hand, depends strongly on the type of TM
ion and reveals a transition in the interaction with the
substrate. As in the $dI/dV$ spectra, the PDOS shows that this
state is unoccupied in FePc, partially occupied in CoPc, and
acquires a charge of approximately one electron in NiPc and CuPc.
The splitting between occupied and unoccupied $2e_g$ states is not fully
reproduced in the calculations due to the well-known
electronic gap problem in DFT, which is
especially critical for delocalized $\pi$-orbitals.~\cite{Yang08S}

The origin of the charge-transfer behavior with TM can be unfolded
by using gas-phase anions to study pure transfer without
hybridization. We find that the TM-$d$ dependent behavior is the
same as in adsorbates. In [FePc]$^{1-}$ and [CoPc]$^{1-}$ the effect of the extra electron is complex and the extra charge is not easily assigned to a single MO. It rather induces a reorganization of the charge within the molecule. In [NiPc]$^{1-}$ and [CuPc]$^{1-}$ the behavior is simpler, with the extra electron going to $2e_g$.
The observed behavior is also in agreement with transport results
in electron-doped thin films.~\cite{Morpurgo05APL}

In summary, although the net amount of transferred charge is about
one electron in all cases (see Table~\ref{tab3}), its effect on the molecule depends on the $d$ occupation of the ion. This difference
is driven by the interplay between energy alignment and Coulomb
repulsion in the $d$ and $2e_g$ states, as well as their degree of hybridization with the substrate. The similarity between
results obtained by calculating gas-phase anions and adsorbates
indicates that the leading role of the substrate is that of a
charge reservoir. However, the strong (weak) hybridization of the
$d_\perp$ ($d_\parallel$) states further stabilizes the different
charge donation channels, promoting charge transfer to TM-$d$
states in FePc and CoPc, and inhibiting it in NiPc and CuPc.

\section{MAGNETIC PROPERTIES}
\label{sec:mag_struc}

The magnetic properties of TMPc on Ag(100) are addressed using STS
through the analysis of $dI/dV$ spectra close to $E_F$. The
presence of sharp zero bias resonances can be used to detect the
formation of local magnetic moments screened by the Kondo
effect.~\cite{Hewson93,Schneider09JoPCM} This occurs as itinerant
electrons belonging to the substrate couple antiferromagnetically
to a localized spin, forming a many-body singlet state at
temperatures below the characteristic Kondo energy $k_BT_K$.
$dI/dV$ maps of the Kondo spectral features further allow us to
measure the spatial distribution of the spin density for unpaired
electrons. We note, however, that, although the presence of a
Kondo resonance is a signature of a local moment, its absence is
not a proof of a nonmagnetic ground state, since the resonance can also be experimentally inaccessible due to the low $T_K$ in weakly interacting spins. In such a case, STS
data have to be contrasted with complementary experiments, such as
x-ray magnetic circular
dichroism,~\cite{Gambardella10PRB,Gambardella11PRB} or, as in the
present case, with theoretical calculations.

\subsection{Kondo effect probed by STS}

\begin{figure}[!h]
\includegraphics[width=\columnwidth]{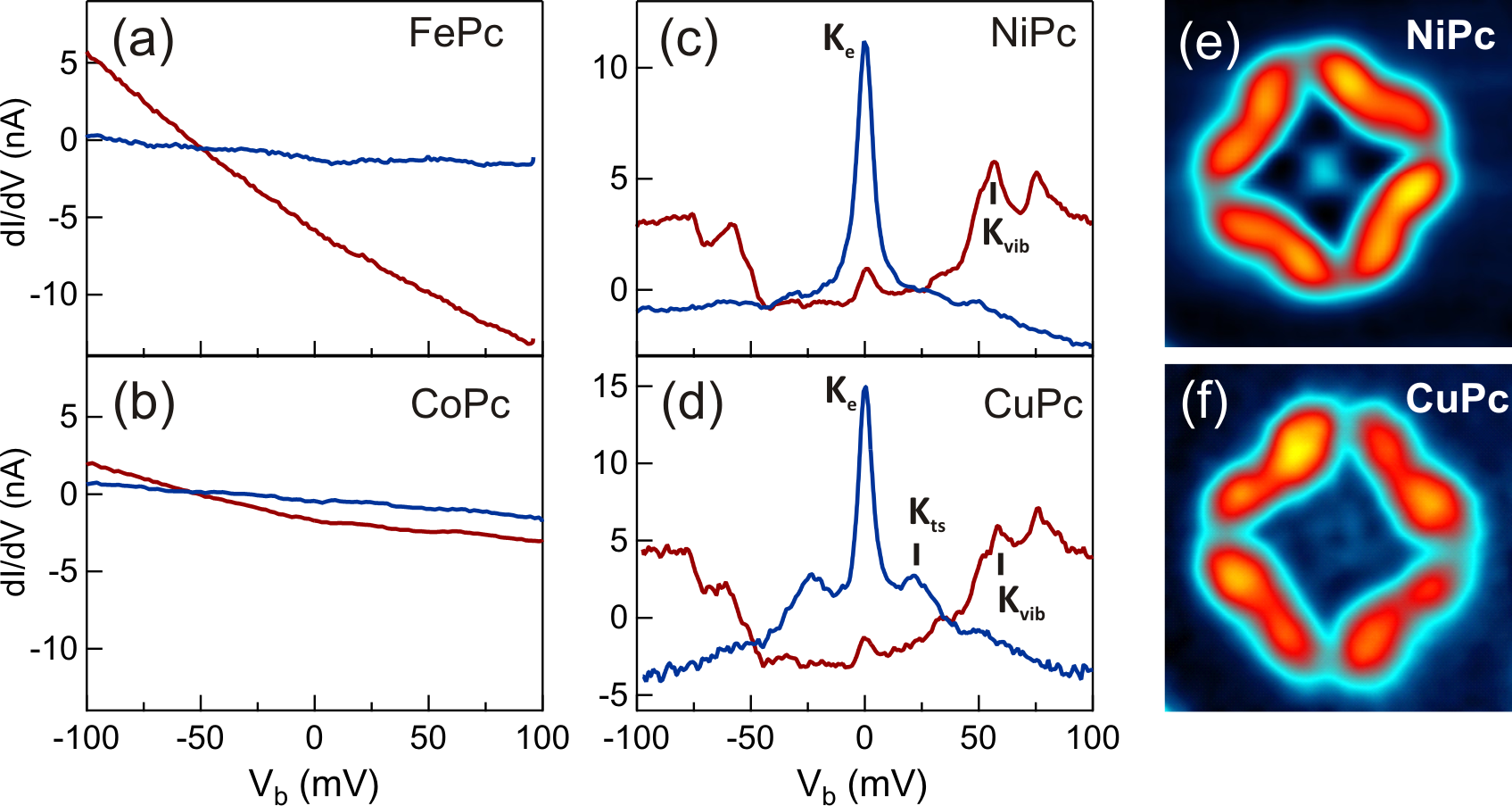}
\caption{(Color online). (a)-(d) Conductance spectra acquired
around $E_F$ on the TM ions (red) and benzene rings (blue). The
zero-bias (elastic), vibrational, and triplet-singlet Kondo
resonances observed in CuPc and NiPc are labelled as $K_e$,
$K_{v}$, and $K_{ts}$ respectively. Tunneling conditions ($I$,
$V_b$):  1 nA,  -100 mV (FePc, CoPc), 1.1 nA, -100 mV (NiPc), 2
nA, -100 mV (CuPc). (e) and (f) $d^2I/dV^2$ maps acquired at -3 mV
and -5mV respectively, showing the intensity distribution of the
elastic Kondo resonance ($K_e$) of CuPc and NiPc and its close
resemblance to the $2e_g$ resonances. \label{FIG_7}}
\end{figure}

Figure~\ref{FIG_7} shows the $dI/dV$ spectra measured near
$E_F$ for the four molecules at the position of the TM ion (red
curves) and benzene ligand (blue curves). FePc and CoPc present
featureless spectra, which indicate that the Kondo interaction
between these molecules and the Ag(100) surface is either absent
or too weak to be observed at 5~K. Relatively flat Co spectra
were reported also for CoPc adsorbed on Au(111), for which it was
concluded that the filling of the $a_{1g}$ state leads to a
complete quenching of the molecular magnetic moment.~\cite{Zhu05S}
This interpretation has been recently questioned based on a
mixed-valence model of x-ray absorption spectra.~\cite{Gambardella11PRB} In both cases, the
quenching of the CoPc magnetic moment appears to be a robust
result. On the other hand, FePc does
present a Kondo resonance on Au(111), although different
interaction strengths have been
reported.~\cite{Gao07PRL,Kawai11PRL} We attribute the absence of
Kondo peaks on Ag(100) to a stronger interaction with the
substrate, as reflected by the substantial hybridization of the
$d_\perp$ states, which may lead to a mixed-valence configuration
of FePc, as discussed later.

\begin{figure}[!h]
\includegraphics[width=\columnwidth]{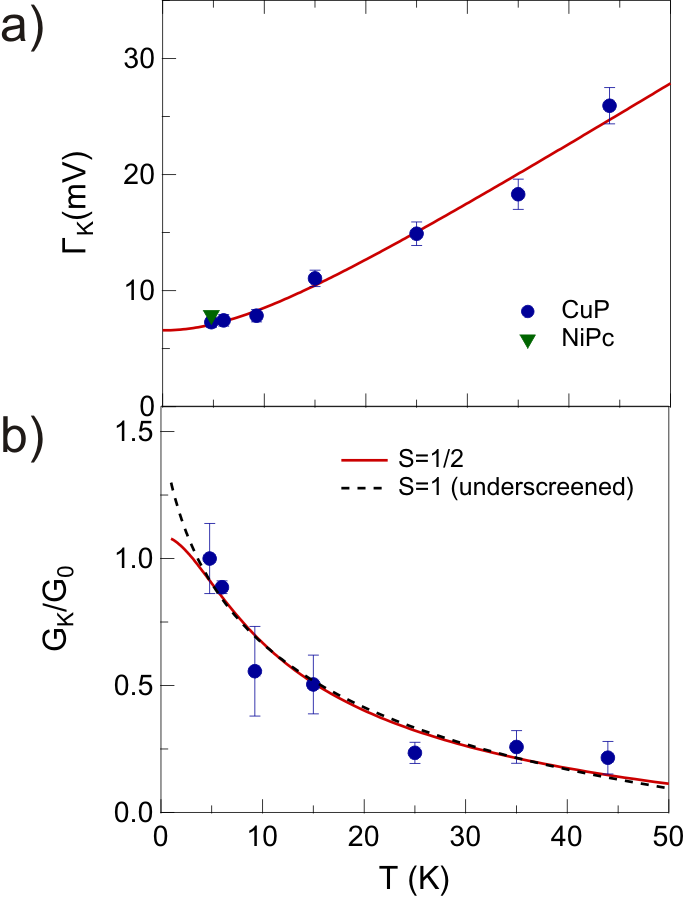}
\caption{(Color online). Temperature dependence of (a) $\Gamma_K$
and (b) $G_K$ normalized to the intensity at 5 K ($G_0$) of the
elastic Kondo resonance of CuPc. The solid line in (a) is a fit of
$\Gamma_K$ using Eq.~\ref{GammavsT} giving $T_K = 27 \pm 2$~K. The
solid (dashed) line in (b) is a fit of $G_K/G_0$ for the S=1/2
(underscreened S=1) case of Eq.~\ref{GvsT}. \label{FIG_8}}
\end{figure}

In contrast with FePc and CoPc, NiPc and CuPc present many
characteristic signatures of low-energy excitations around $E_F$.
Such excitations have been discussed in detail in
Ref.~\onlinecite{Gambardella11NC}. Here we report a summary of these
results for comparison with theory as well as to provide a
complete picture of TMPc adsorbed on Ag(100). The most prominent
feature of the NiPc and CuPc spectra is the presence of an intense
zero bias resonance with maximum intensity at the ligand position.
This is rather unusual for metal-organic adsorbates since, in most
cases, Kondo resonances are associated to the spin of the TM
ions.~\cite{Zhu05S,Xue07PRL,Xue08PRL,Gao07PRL,Kawai11PRL} Here, the zero bias peak is assigned to an elastic
Kondo resonance ($K_{e}$) arising from the unpaired spin located
in the $2e_g$ orbital of CuPc and NiPc. This interpretation is
supported by the temperature dependence of the peak´s full width at
half maximum ($\Gamma_K$) and maximum intensity ($G_K$).
Figure~\ref{FIG_8} shows that both parameters follow
expressions derived from Fermi liquid theory~\cite{Crommie02PRL}
and numerical renormalization group (NRG):~\cite{Ralph10S}

\begin{equation}
\Gamma_K(T)= 2 \sqrt{(\pi k_BT)^2+2(k_BT_K)^2},
\label{GammavsT}
\end{equation}

\begin{equation}
 G_K(T)=G_{off}+G(0) \cdot \left[1+\left(\frac{T}{T_K}\right)^\xi \cdot (2^{1/\alpha}-1)\right]^{-\alpha}.
\label{GvsT}
\end{equation}
where $\alpha$ and $\xi$ are parameters that depend on the total
spin of the molecule. Fitting the values of $\Gamma_K$ with Eq.~\ref{GammavsT} leads to a  Kondo temperature of $T_K = 27 \pm 2$~K
for CuPc. The logarithmic behavior of $G_K$ makes it more
sensitive to the scattering of the data, but we find that
expressions for both S=1/2 and (the underscreened) S=1 fit
reasonably well the data by using $T_K = 27$~K. The two possible
spin states will be discussed later.

We notice that NiPc and CuPc present very similar Kondo resonances
in spite of the S=0 ground state of the Ni ion and S=1/2 ground
state of Cu. Indeed, we estimate a Kondo temperature $T_K = 29$~K
for NiPc from $\Gamma_K$ measured at 5 K. Moreover, the
intensity of both CuPc and NiPc resonances closely follows the
contour of the $2e_g$ orbital, as revealed by the remarkable
similarity of the $dI/dV$ maps reported in
Figs.~\ref{FIG_7}(e) and (f) compared to Fig.~\ref{FIG_5}. These observations indicate the presence of an
unpaired spin localized on the $2e_g$ ligand state, in agreement
with the single occupancy of this orbital discussed in
Section~\ref{sec:elec_struc}. The relation between energy, width,
and Coulomb repulsion potential, with values of $\epsilon= -0.30$
eV, $\Gamma\sim0.30$ eV,  and $U=0.65$ eV, respectively, is also
characteristic of the Kondo regime, for which $|\epsilon|,
|\epsilon+U|\geq \Gamma$.

The direct observation of the "Coulomb blockade" peaks and the
Kondo resonance allows, unlike in previous studies of molecular
adsorbates, for a univocal identification of the MO associated to
the unpaired spin. As the spin participating to the many-body
Kondo state originates from a single orbital, we expect a
well-defined value of $T_K$ over the whole molecule, as confirmed
by our observations. This is in contrast with the variations of
$T_K$ found for Co porphyrins adsorbed on Cu(111), which suggest a
spatially-dependent, multiple orbital origin of the Kondo
interaction, possibly due to strong intermixing of TM and ligand
orbitals near $E_F$ (Ref.~\onlinecite{Hla10PRL}).

\subsection{Inelastic vibrational and Kondo excitations}

\begin{figure}[!t]
\includegraphics[width=\columnwidth]{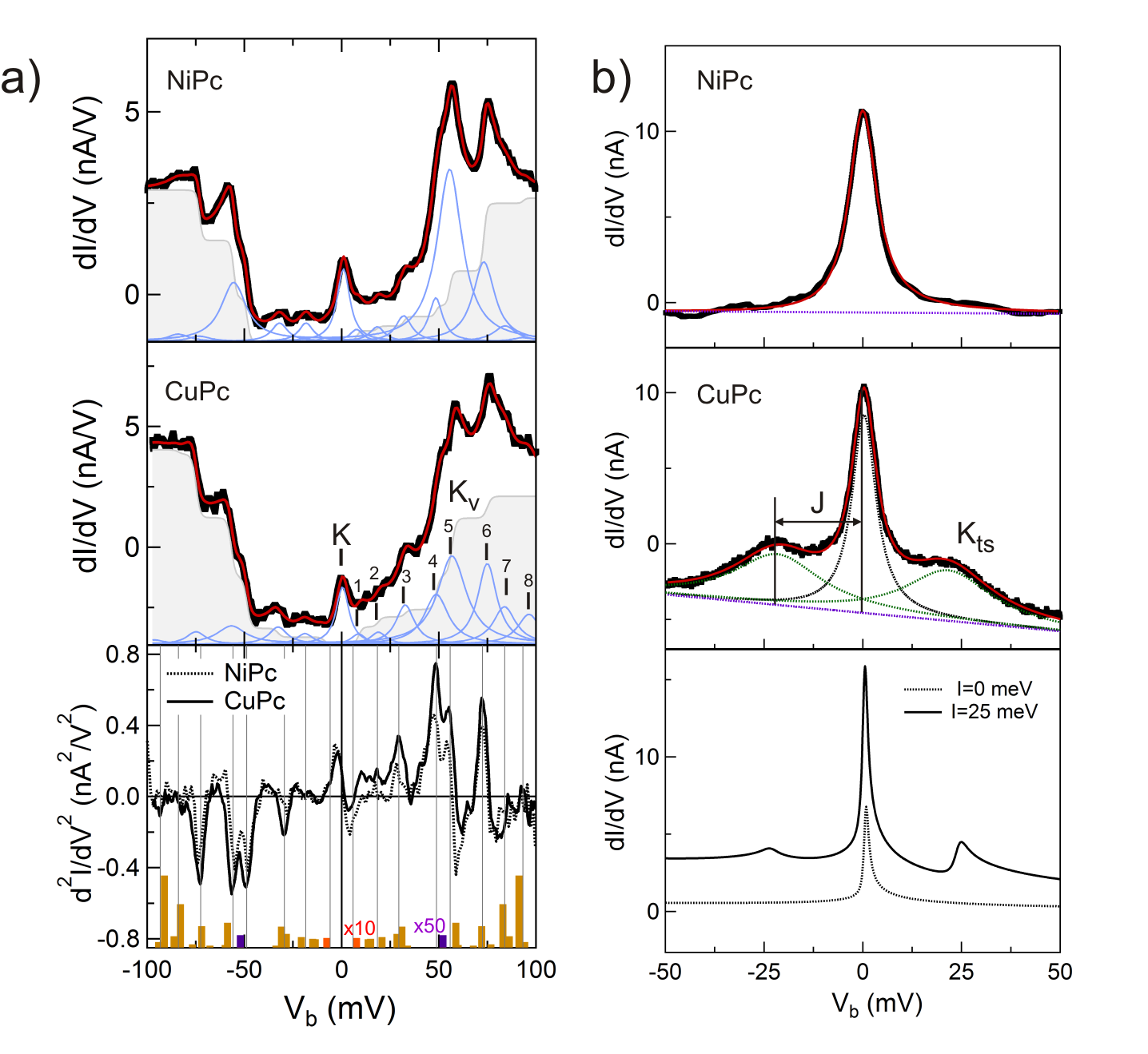}
\caption{(Color online).  Fit of the differential conductance
($dI/dV$) spectra of NiPc and CuPc measured at (a) the position of
the TM ion and (b) the benzene ring, using Fano  and Lorentzian
functions for the zero and finite bias peaks, respectively, and
step functions for the inelastic conductance steps. The bottom
graph in (a) shows the second derivative ($d^2I/dV^2$) of the
spectra measured on Ni and Cu. The energies obtained from the fits
are indicated by dashed lines in the $d^2I/dV^2$ spectra. The bars
correspond to the intensity of calculated Raman (yellow) and
infrared (grey) modes of gas-phase CuPc.~\cite{Zhou05VS} The bottom
graph in (b) shows the theoretical PDOS obtained for the $2e_g$
orbital of CuPc for $I$ = 25 meV (solid line) and 0 meV (dashed
line) calculated within the non-crossing approximation (see text
for details). \label{FIG_9}}
\end{figure}

The $dI/dV$ spectra of NiPc and CuPc recorded on the TM ions [red
curves in Figs.~\ref{FIG_7}(c) and (d)] present a complex
series of peaks and conductance steps in addition to the elastic
Kondo resonance observed at zero bias. It is well-known that
inelastic excitations induce step-like increases in the
differential conductance spectra symmetrically distributed in
energy around $E_F$.~\cite{Ho98S,Gauyacq09PRL} In Kondo systems,
the coherent coupling of the Kondo state with such excitations
leads to additional peaks at the energies of the
inelastic conductance steps.~\cite{Kiselev07pssc} Unlike quantum
dots, where the reduced orbital level-spacing favors the coupling
of Kondo and electronic excitations,~\cite{Rogge10NL} the origin of
such side peaks in small aromatic molecules is restrained to
vibrational~\cite{Flensberg05PRL,Natelson04PRL,Gupta10NL,Pascual08PRL}
($K_{v}$) or
magnetic~\cite{Shtrikman03PRB,Bjornholm10NL,Nygaard06NP,Ralph10S,Balestro08N}
($K_{ts}$) excitations. In the present case, by comparing the
energy of the $K_{v}$ peaks with that of the Raman vibrational
modes of gas-phase CuPc,~\cite{Zhou05VS} we find that the inelastic
features found at the position of the Ni and Cu ions
[Fig.~\ref{FIG_9}(a)] correspond to vibrational
excitations that involve distortion of the TM-N$_p$ bonds.~\cite{Gambardella11NC} Note that the
vibrational spectra of NiPc and CuPc are nearly identical, as shown by
the twin $d^2I/dV^2$ spectra displayed in the bottom graph. The
step-like and peak-like contributions to the inelastic conductance
can be obtained by fitting the $dI/dV$ spectra with step and
Lorentzian functions, as shown in Fig.~\ref{FIG_9}(a).
From the fit of the spectra we observe that both NiPc and CuPc
present more intense conductance steps at negative bias and more
intense Kondo peaks at positive bias. These inverted correlation
between vibrational and Kondo features has also been observed for
TCNQ molecules adsorbed on Au(111), and has been attributed to the
competition between the purely inelastic channel and the one
including the Kondo effect.~\cite{Pascual08PRL}

\subsection{Inelastic triplet-singlet and Kondo excitations}

In contrast to the spectra of the TM ions, the spectrum of NiPc
recorded on benzene does not present signatures of finite-bias
excitations. The CuPc benzene spectrum, however, presents two
side-peaks at $\pm 21 \pm 1$~meV, shown in
Fig.~\ref{FIG_9}(b). These side-peaks are assigned to
intramolecular magnetic excitations,~\cite{Gambardella11NC} owing to
the presence of two spins in CuPc, one localized on the Cu
$d_{x^2-y^2}$ state,~\cite{Gambardella10PRB} and the other in the
$2e_g$ ligand state. This interpretation is supported by the
gas-phase calculation of charged molecules presented later. The
energy of the side peaks provides a direct estimate of the
intramolecular exchange coupling ($J$) between $d$- and
$\pi$-spins. The observation of an intense Kondo peak at zero
bias, on the other hand, indicates that the CuPc ground state is
magnetic, namely a triplet (S=1) state where the metal and ligand
spins are aligned parallel to each other, whereas the excited
state is a singlet (S=0) state.

The similarity between the elastic Kondo resonances of NiPc and
CuPc, with S=1/2 and S=1 respectively, can be explained by the
fact that the total CuPc spin is underscreened due to the very
weak coupling between the $b_{1g}$ ($d_{x^2-y^2}$) orbital and the
substrate. Thus, only the $2e_{g}$ ligand spin is effectively
screened by the substrate in the temperature range accessible by
our STM.

\subsection{Multi-orbital non-crossing approximation of the Kondo resonance associated to the ligand spin}

Although spin excitations coupled to the Kondo effect have already
been observed in transport experiments with molecular
junctions,~\cite{Shtrikman03PRB,Bjornholm10NL,Nygaard06NP,Ralph10S,Balestro08N}
the precise configuration of the molecule-metal bonding provided
by STM and the assignment of the Kondo-screened spin to a specific
orbital allow us to model our results by mapping the electronic
structure of CuPc/Ag(100) calculated by DFT onto an impurity
Hamiltonian. Our model describes the interaction between the
$2e_g$ orbital and the substrate by an SU(4) Anderson Hamiltonian,
where two-fold spin degeneracy coexists with the two-fold orbital
degeneracy on the same footing, and an infinite repulsive Coulomb
term (see Ref.~\onlinecite{Lorente11JoPCM} for more details). DFT
calculations show that the interaction between the ion $b_{1g}$
state and the substrate can be neglected, as reflected by the
small hybridization-induced broadening obtained for this orbital,
which is a factor of 20 smaller than that of the $2e_g$ state. The
Hamiltonian includes a molecular exchange term $-J S_L \cdot
S_{TM}$, which couples the ligand spin ($S_L$) to the TM spin
($S_{TM}$). The $2e_g$ on-site energy is fitted to $\epsilon$ = -0.35 eV,
and the spectral density of the coupling with the substrate is
assumed to have rectangular shape with magnitude $\Gamma$ = 0.1 eV
and band half-width of 10 eV.

The PDOS onto the $2e_g$ state, calculated within the non-crossing
approximation at T=5 K (see Ref.~\onlinecite{Lorente11JoPCM}), is
compared with the experimental $dI/dV$ spectra in
Fig.~\ref{FIG_9}(b). For STM studies this comparison
is a first-order approximation that takes into account that the
tunneling conductance is proportional to the local DOS. Moreover, our
calculations are performed in equilibrium. At typical STM
parameters, the non-equilibrium modification of NCA
equations~\cite{Hershfield98PRB} for slowly varying substrate DOS,
reduce to corrections smaller than 1/1000$^{th}$ in the tunneling
current. Hence, in these STM studies, the equilibrium spectral
function is an excellent approximation to compare with the
experiment. This is confirmed by the correct reproduction of the
Kondo features observed for NiPc and CuPc when we set the
intramolecular coupling term to 0 meV and 25 meV respectively, as
shown in the bottom graph of Fig.~\ref{FIG_9}(b). In
the case of ferromagnetic coupling, the finite-bias features
correspond to inelastic Kondo replicas that originate from spin
excitations. Decomposition into spin channels yields that the
Stokes satellite (positive bias) is in the S = 0 channel, whereas
the anti-Stokes peak (negative bias) as well as the Kondo peak are
in the S = 1 channel.~\cite{Lorente11JoPCM,Aligia09PRB}

\subsection{Magnetic moments calculated by DFT}

\begin{figure}[!h]
\includegraphics[width=\columnwidth]{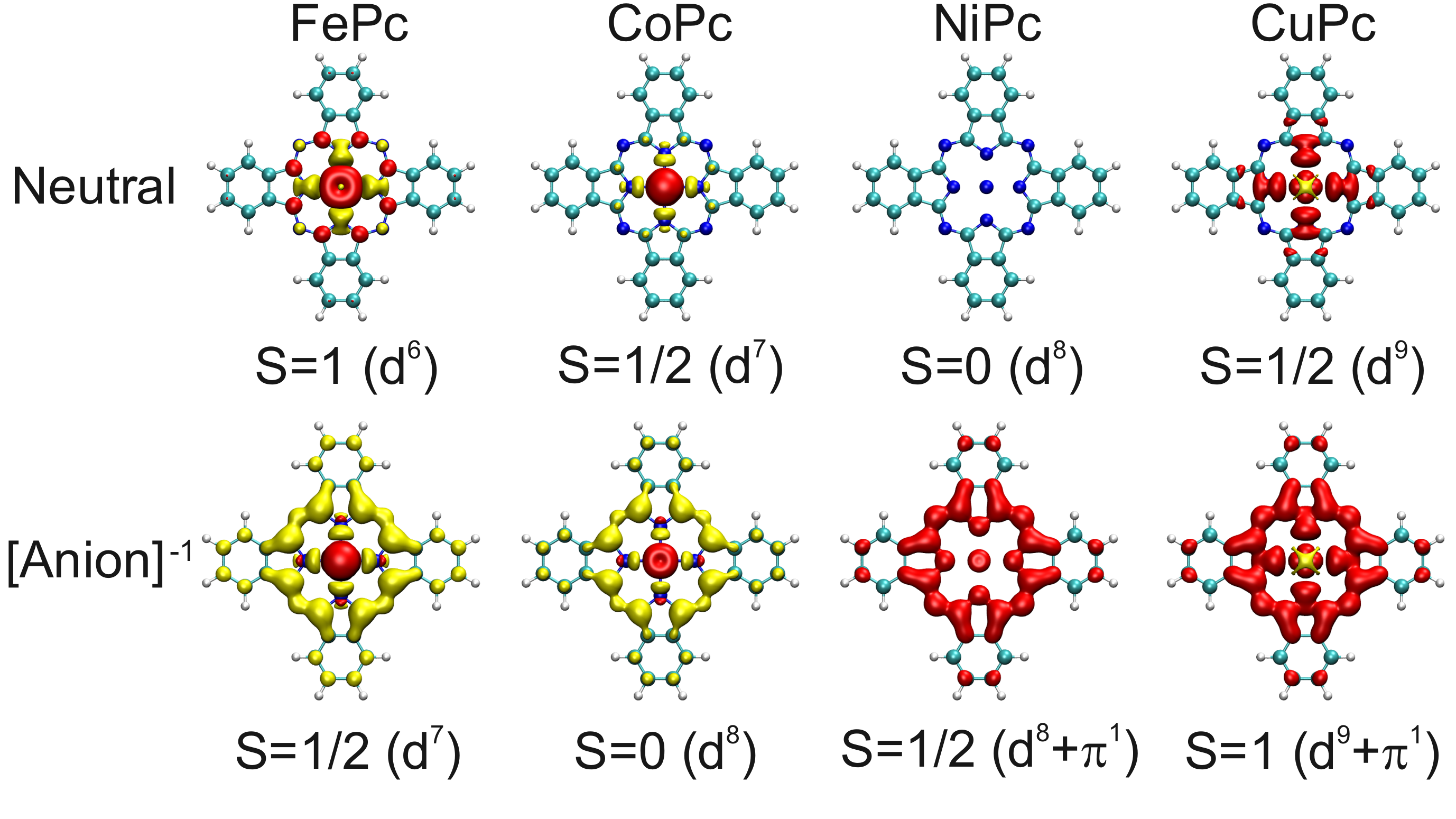}
\caption{(Color online). Spin density of neutral (top) and anionic
(bottom) gas-phase TMPc in $D_{4h}$ symmetry. Red (yellow)
indicate spin up (down). The isovalue for the spin density is
0.005~$e^-$/\AA$^3$. \label{FIG_10}}
\end{figure}

By calculating the saturation magnetic moments of TMPc, DFT
provides complementary results to the analysis of the Kondo
spectra reported above. Moreover, in addition to supporting the
interpretation of the STS data, DFT allows us to construct a
quantitative picture of the effect of the different charge
transfer mechanisms on the molecular magnetic properties.
Table~\ref{tab3} summarizes the magnetic moments calculated for
the neutral and anionic form of gas-phase TMPc as well as for TMPc
adsorbed on Ag(100). We notice that, although the amount of charge
transferred from the substrate to the molecules is similar in all
cases, the total magnetic moment is reduced in FePc and CoPc,
whereas it is increased in NiPc and CuPc with respect to the
neutral gas-phase TMPCs. The calculations confirm that the
magnetic moment of the TM ions in NiPc and CuPc is not perturbed
upon adsorption, owing to the small hybridization of the planar
$b_{1g}$ orbital with the substrate states. This is in agreement
with previous results obtained by XMCD on CuPc/Ag(100)
(Ref.~\onlinecite{Gambardella10PRB}) and with the coexistence of
TM and ligand spins in CuPc inferred from the Kondo spectra. We
recall that, due to the underestimation of correlation effects in
DFT, the ligand spin is just barely present in the calculations
with adsorbates. For NiPc, we calculate a small magnetic moment of
0.14$\mu_B$, while for CuPc the moment is just slightly larger
than the 1$\mu_B$ corresponding to the unpaired spin in the
$b_{1g}$ orbital. Since charge transfer is close to one electron,
one way to estimate the magnitude of the ligand spin using DFT is
to consider gas-phase anions as a proxy of adsorbed TMPc. In such
a case the number of electrons belonging to the molecule is fixed
and the calculations can be constrained to yield the minimum energy
spin state. Figure~\ref{FIG_10}
displays the spin density of neutral and anionic gas-phase
molecules, where we clearly observe the additional ligand spin of
NiPc and CuPc upon charge transfer. In CuPc, the spin density
originating from the $b_{1g}$ orbital is also observed,
distributed over the Cu ion and N$_p$ atoms.

\begin{table}[!h]
\centering

\begin{tabular}{c|c|ccc}
         \hline
        \hline
          & $\Delta N$ &  $m_{TMPc}^{gas} $ &  $m_{[TMPc]^{1-}}^{gas} $ & $m_{TMPc}^{ads}$\\
        \hline
        FePc  & 0.80 & 2.00 & 1.00 & 1.06  \\
        CoPc  & 0.99 & 1.00 & 0.00 & 0.63  \\
        \hline
        NiPc  & 1.13 & 0.00 & 1.00 & 0.14  \\
        CuPc  & 0.81 & 1.00 & 2.00 & 1.32  \\
        \hline
        \hline
\end{tabular}
\centering \caption[] {\label{tab3} Computed charge transfer
$\Delta N$  (in electrons) for the adsorbed molecules, and total
magnetic moment $m$ (units of $\mu_B$) in neutral and anionic
gas-phase, and adsorbed molecules.
}
\end{table}

\begin{figure}[!h]
\includegraphics[width=\columnwidth]{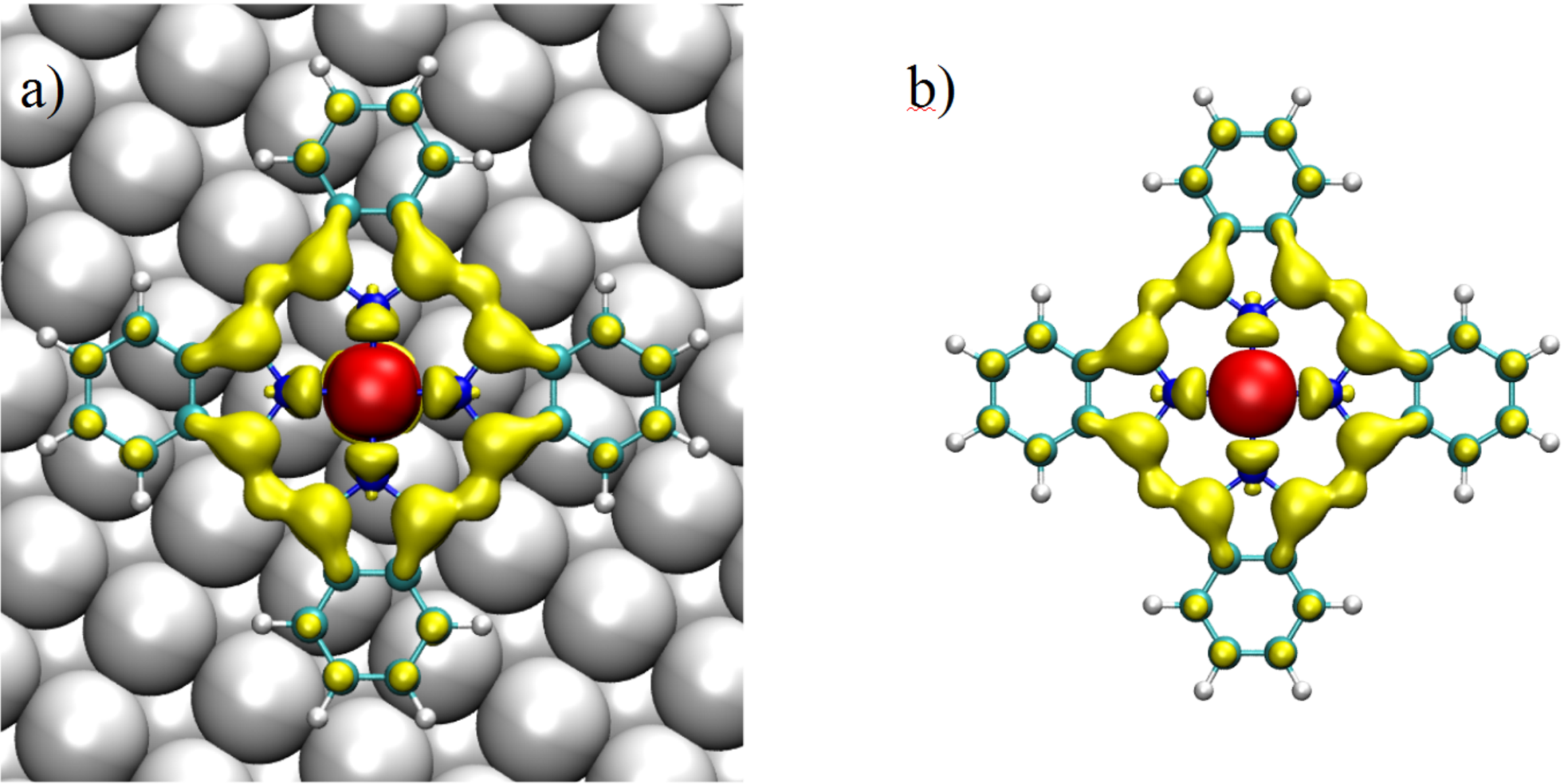}
\caption{(Color online). Spin density distribution of FePc (a)
adsorbed on Ag(100) and (b) in the gas-phase anion. Red/yellow
indicate spin up/down. The isovalue for the spin density is
0.005~$e^-$/\AA$^3$ in both cases. \label{FIG_11}}
\end{figure}

FePc and CoPc present a more complicate picture compared to the
simple addition of one spin in NiPc and CuPc, partly because of
the strong hybridization with the substrate discussed in
Section~\ref{sec:PDOS}. In both cases, the magnetic moment is
reduced upon adsorption. The extra electron donated by the surface
is not easily assigned to a single MO, inducing a reorganization
of the charge within the molecule. We use the results obtained for
gas-phase anions to disentangle the different mechanisms
responsible for the quenching of the magnetic moment. As shown in
Fig.~\ref{FIG_11}, the spin density of FePc follows
the same distribution in the gas-phase anion and in the adsorbed
molecule. This similarity can also be checked by comparing their
element-resolved magnetic moments in Table~\ref{tab4}, which are very
similar in both cases. Thus, regardless of the strong
hybridization of adsorbed FePc, the main features of the spin
moment distribution are again captured by the simple addition of
one electron to the gas-phase system. The true nature of the
ground state of FePc on Ag (100), however, might be more complex
than suggested by the calculated spin density distribution, due to
the interplay of several MO close to $E_F$ and dynamic electron
correlation effects that are not included in DFT. The computed
electronic structure for FePc yields a S=1/2 system when adsorbed
on Ag (100), same as for the gas-phase anion. However, the adsorbate exhibits a more complex configuration due to the presence of multiple molecular orbitals at the Fermi level. This could result in either a mixed-valence system,
given the Fermi level crossing and strong interaction of the
$a_{1g}$ orbital with the substrate, or a more complex Kondo
behavior with a low-temperature Kondo phase originating in the
many less-coupled orbitals near the Fermi energy. This complex
behavior makes FePc/Ag(100) an interesting candidate to study very
low temperature Kondo physics in a multi-orbital system.

CoPc represents another case of nontrivial charge transfer
occurring upon adsorption. For example, the $a_{1g}$ orbital is
filled, but new empty $d_{\pi}$ states appear above $E_F$, as
shown in Figs.~\ref{FIG_4}(b) and (c). Moreover, the anion
is not a good model for the adsorbed molecule. As shown in
Table~\ref{tab3} and Fig.~\ref{FIG_10}, the
anion is in a S=0 state, due to antiferromagnetic coupling between
spins residing in different MOs, while in the adsorbed case we
find a non-integer spin magnetic moment of 0.63$\mu_B$. This
non-integer spin, together with the absence of a Kondo
interaction, could indicate that CoPc is in the mixed-valence
regime, with charge fluctuating between the CoPc $d$ and Ag
states, similar to results obtained for CoPc on Au(111) using
x-ray absorption spectroscopy.~\cite{Gambardella11PRB}

\begin{table}[!h]
\centering

\begin{tabular}{lcccc}
      \hline
        \hline
          & $m_M$ &  $m_N$ & $m_C$ & $m_{tot}$    \\
        \hline
        FePc$_{gas}$         & 1.94 & -0.13 &  0.19 & 2.00  \\
     $[$FePc$]^{1-}_{gas}$   & 2.04 & -0.31 & -0.73 & 1.00  \\
        FePc$_{Ag(100)}$     & 2.14 & -0.26 & -0.82 & 1.06  \\

        \hline
        \hline
\end{tabular}

\centering
\caption[] {\label{tab4} Computed  element-resolved magnetic moment $m$
(units of $\mu_B$) for FePc in the neutral and anionic gas-phase, and in the adsorbate,
using GGA+U.}
\end{table}

\subsection{Intramolecular exchange coupling}

The TM-ligand spin coupling observed via the inelastic Kondo
interaction in CuPc can be studied in gas-phase [CuPc]$^{1-}$.
Here, the energy of ferromagnetic and antiferromagnetic spin
configurations can be computed by fixing the total spin of the
molecule. The ferromagnetic (triplet) alignment is favored over
the antiferromagnetic (singlet) one. The triplet
ground state is in agreement with the zero-bias Kondo resonance
observed experimentally. From the energy difference between
triplet and singlet states, evaluated using the
parallel spin ($E_{\uparrow \uparrow}$) and antiparallel spin ($E_{\uparrow \downarrow}$) energies, and taking the
correct singlet and triplet configurations to yield $J=2 (E_{\uparrow \downarrow} - E_{\uparrow \uparrow})$,
we extract the exchange coupling
constant $J$ for the anion, $J=38$~meV. The discrepancy with
$J=21$~meV measured for CuPc/Ag(100) can be explained by surface
screening, which is expected to significantly reduce exchange
correlation effects involving hybridized orbitals.

The distribution of the spin density in [FePc]$^{1-}$ and
[CoPc]$^{1-}$, shown in Fig.~\ref{FIG_10},
reveals a different type of intramolecular spin correlation. Both
neutral and anionic molecules exhibit antiferromagnetically
aligned TM and ligand spins, indicating that the spin contrast is
not directly due to the interaction with the substrate, but is
intrinsic to the molecule and related to the pristine TM spin,
which extends to nearby C and N atoms upon the formation of MOs
with mixed $d$- and $\pi$-character. Such intramolecular
antiferromagnetic spin coupling is a result of the large exchange
splitting of the $d$-levels of the Fe and Co ions. This
induces spin-dependent mixing of the $d$ and ligand states,
leading to the formation of spin-polarized MOs with different
spatial distribution. A similar mechanism can be invoked to
explain the site- and energy-dependent spin contrast recently
measured for CoPc deposited on Fe(110), for which the total
molecular moment is zero.~\cite{Wiesendanger10PRL}

\section{CONCLUSIONS}
\label{sec:Conclusions}

The adsorption, electronic, and magnetic structure of four TMPc (M
= Fe, Co, Ni, and Cu) adsorbed on Ag(100) have been systematically
studied using STM, STS, and {\it ab-initio} DFT calculations
within the LDA, GGA, GGA+vdW, and GGA+U approximations.

We find that all molecules adsorb with the macrocycle plane
parallel to the surface and the TM-N$_p$ ligand axis rotated by
$\pm$30$^\circ$ with respect to the high-symmetry surface lattice
vectors. The comparison between DFT results using different
approximations evidences differences in the TM-Ag interaction, but
the leading role of vdW forces leads to similar molecule-substrate
distances and azimuthal rotation angles. The rotation between
molecular and surface symmetry axes imprints chirality to the
frontier electronic orbitals without perturbing the structural
conformation of the molecules. This can only be observed in NiPc
and CuPc at negative voltages, due to the contribution of the
$a_{1u}$ and partially occupied $2e_g$ orbitals to the tunneling
current and minor influence of the $d$-channels in the same energy
region.

The electronic structure of adsorbed TMPc is modified mainly due
to charge transfer from the substrate. Although all molecules
receive approximately one electron, charge is transferred to
the ligand $2e_g$ orbital in NiPc and CuPc, and to multiple MOs in FePc and CoPc, inducing internal charge reorganization. This difference is related to both energy level
alignment and Coulomb repulsion. In addition, whereas FePc and
CoPc exhibit strongly hybridized $d_\perp$-states near E$_F$,
those are shifted down in energy in NiPc and CuPc, which display
unperturbed $d_\parallel$-states in this energy range. The
electronic configuration of the Ni and Cu ions is not
significantly affected by adsorption.

Both charge transfer and hybridization tend to reduce the magnetic
moment of FePc and CoPc. The computed electronic structure of FePc
indicates a S=1/2 system when adsorbed on Ag(100), with a strong
presence of MOs near the Fermi energy that could indicate either a
mixed-valence system, given the strong interaction of the $a_{1g}$
orbital with the substrate, or a more complex Kondo behavior with
a low-temperature Kondo phase originating in the many less-coupled
orbitals near the Fermi energy. The non-integer moment obtained for CoPc and the
existence of a substantial molecular DOS at the Fermi energy (see Fig.~\ref{FIG_4}), together with the
absence of a Kondo resonance in this molecule, suggest a
mixed-valence configuration. Opposite to the moment reduction in
FePc and CoPc, the spin multiplicity is actually enhanced by
charge transfer in NiPc and CuPc, due to the induction of a ligand
spin in the $2e_g$ orbital. This transforms NiPc into a
paramagnetic molecule and induces a triplet ground state in CuPc,
where the ligand spin is exchange-coupled to the TM magnetic
moment.

The interaction between the $2e_g$ ligand spin of NiPc and CuPc
and the substrate gives rise to a Kondo interaction, which induces
a prominent zero bias resonance delocalized over the perimeter of
the molecules. Coherent coupling between Kondo spin flip and
vibrational excitations induces inelastic Kondo resonances in both
NiPc and CuPc. In CuPc, the total magnetic moment is underscreened
as the Kondo energy scale of the TM spin is orders of magnitude
smaller than the one of the $2e_g$ ligand spin. However, the
additional magnetic degree of freedom due to the TM spin leads to
the coupling of Kondo and inelastic triplet-singlet excitations.
The coexistence of different nonequilibrium Kondo processes
related to vibrational and spin transitions in CuPc opens up the
possibility to study the timescale and spatial localization of
multiple spin excitation and relaxation channels within a single
molecule, as reported in more detail in
Ref.~\onlinecite{Gambardella11NC}.

In general, we show that charge transfer and hybridization of MOs
orbitals have profound effects on the electronic configuration,
magnetic moment, and transport properties of metal-organic
complexes adsorbed on a metallic substrate. Such effects in TMPc
do not depend monotonically on the electronic configuration of the
TM ions. Rather, the energy position, symmetry, and spin
polarization of the pristine MOs has to be considered within a
comprehensive picture of the charge transfer process. Importantly,
the magnetic moment of adsorbed TMPc can be either reduced or
enhanced depending on the relative energy of the $d_\perp$ and
$\pi$-levels, and their degree of hybridization with the substrate. Further, the possibility of inducing additional
ligand spins upon adsorption may be relevant to establish magnetic
coupling in extended molecular structures assembled on metal
surfaces.

\begin{acknowledgements}
We acknowledge support from the European Research Council (StG
203239 NOMAD), Ministerio de Ciencia e Innovaci\'{o}n
(MAT2010-15659), and Ag\`{e}ncia de Gesti\'{o} d'Ajuts
Universitaris i de Recerca (2009 SGR 695). A.M. acknowledges
funding from the Ramon y Cajal Fellowship program. R.R. is
supported by a JAE-Doc contract from the Consejo Superior de
Investigaciones Cient\'{\i}ficas. R.R., R.K and N.L. have been supported by the ICT-FET Integrated Project AtMol (http://www.atmol.eu).
\end{acknowledgements}

\bibliographystyle{apsrev}

\bibliography{MPc_elec_struc_noURL}

\end{document}